\documentclass[namedreferences]{SolarPhysics}
\usepackage[optionalrh,solaenum]{spr-sola-addons} 
\usepackage{subfig}
\usepackage{caption}
\usepackage{multirow}
\usepackage{graphicx}        
 \usepackage{color}           
\usepackage{url}             
 \usepackage{amssymb}          
 
\begin{document}
\begin{article}
\begin{opening}
\title{Coronal Magnetic Field Structure and Evolution for Flaring AR 11117 and its Surroundings}
\author{Tilaye~\surname{Tadesse}$^{1}$\,$^{2}$,
         T.~\surname{Wiegelmann}$^{1}$, and
        B.~\surname{Inhester}$^{1}$, and
        A.~\surname{Pevtsov}$^{3}$
       }
\runningauthor{T.~Tadesse et al.}
\runningtitle{Coronal Magnetic Field Structure and Evolution}

   \institute{$^{1}$ Max Planck Institut f\"{u}r Sonnensystemforschung, Max-Planck Str. 2, D--37191 Katlenburg-Lindau, Germany
                     email: \url{tilaye.tadesse@gmail.com}, email: \url{wiegelmann@mps.mpg.de}, email: \url{inhester@mps.mpg.de}\\
                $^{2}$Addis Ababa University, College of Natural Sciences, Institute of Geophysics, Space Science, and Astronomy,
                Po.Box 1176, Addis Ababa, Ethiopia\\
              $^{3}$ National Solar Observatory, Sunspot, NM 88349, U.S.A.
                     email: \url{apevtsov@nso.edu} \\
             }

\begin{abstract}
In this study, photospheric vector magnetograms obtained with the \textit{Synoptic
Optical Long-term Investigations of the Sun survey} (SOLIS), are used as boundary
conditions to model the three-dimensional nonlinear force-free (NLFF) coronal
magnetic fields as a sequence of nonlinear force-free equilibria in spherical geometry.
We study the coronal magnetic field structure inside active regions and its temporal evolution.
We compare the magnetic field configuration obtained from NLFF extrapolation before and after
flaring event in active region (AR) 11117 and its surroundings observed on 27 October 2010.
We compare the magnetic field topologies and the magnetic energy densities and study
the connectivities between AR 11117 and its surroundings. During the
investigated time period, we estimate the change in free magnetic energy from before to after
the flare to be $1.74\times 10^{32}\,\textrm{erg}$ which represents about $13.5\%$ of nonlinear
force-free magnetic energy before the flare. In this study, we find that electric currents from
AR 11117 to its surroundings were disrupted after the flare.

\end{abstract}
\keywords{Solar flare · Magnetic fields · Photosphere · Corona}
\end{opening}
\section{Introduction}
     \label{S-Introduction}
The structure of the Sun's corona is dominated by its magnetic field. To understand
eruptive events (flares, coronal mass ejections (CMEs) or filament eruptions), we need
to follow the evolution of the 3D magnetic configuration (geometry and topology) \cite{Schrijver:2011}.
Knowledge of the amount of free magnetic energy and its temporal variation
during CMEs/flares will help our quantitative understanding of solar explosive phenomena
\cite{Bleybel:2002,Regnier:2006,Thalmann,Jing:2010}. However, routine measurements of the solar
magnetic field are mainly carried out in the photosphere. The difficulties of measuring
the coronal field and its embedded electrical currents thus leads us to use
numerical modelling to infer the field strength in the higher layers
of the solar atmosphere from the measured photospheric field.

Nonlinear force-free field (NLFFF) models are thought to be viable tools for investigating
the structure, dynamics, and evolution of the coronae of solar active regions. NLFFF models were
successfully applied to analytical test cases \cite{Schrijver06,Metcalf:2008}, but
they were less successful in applications to real solar data. Different NLFFF models have been
found to yield markedly different field line configurations and widely varying estimates
of the magnetic free energy in the coronal volume, when applied to solar data \cite{DeRosa}. The main
reasons for that problem are (1) the forces acting on the plasma within the photosphere, (2) the uncertainties
of vector-field measurements, particularly of the transverse component, and (3) the large domain that
needs to be modelled to capture the connections of an active region to its surroundings. Therefore,
nonlinear force-free modeling is not a routine procedure which is guaranteed to produce meaningful
results unless the above points are taken into account \cite{DeRosa,Schrijver:2009}. In this study, we
have considered those points explicitly. However, caution is still needed when assessing results
from this modeling. This is because many aspects of the specific approach to modeling
used in this work, such as the use of preprocessed boundary data, the missing boundary data,
measurement error due to noise, resolution of a magnetogram and the departure of the model
fields from the observed boundary fields may influence the results.

Solar flares are thought to be powered by the magnetic free energy (i.e., the difference
between the actual magnetic energy and energy of the equivalent potential field) stored in the
corona prior to eruption. The storage of free energy requires a non-potential magnetic field,
and is therefore associated with a shear or twist in the coronal field away from the potential,
current-free state \cite{Priest:2002,Su:2007,Murray:2011}. To date, the pre-cursors to flaring are still
not fully understood, although there is evidence that flaring is related to changes in the
topology or complexity of an active region's magnetic field.

In this study, we model the coronal magnetic field to determine the sources of flaring activity
and the temporal evolution of an active region between pre- and post-flare stages. Assuming that the evolution
of the coronal magnetic field above an active region can be described by successive equilibria,
we follow in time the magnetic changes of the 3D nonlinear force-free (nlff) fields reconstructed
from two photospheric vector magnetograms taken before and after a flare. We use photospheric
vector magnetograms as the boundary conditions to model the three-dimensional coronal magnetic fields
in spherical geometry. This enables us to accommodate most of the connectivities within
AR 11117 and its surroundings.
\section{Nonlinear force-free field extrapolation}
Except for during eruptions, the magnetic field in the solar corona evolves slowly as it responds
to changes in the surface field, implying that the electromagnetic Lorentz forces in this
low-$\beta$ environment are relatively weak and that any electrical currents that exist must
be essentially parallel or antiparallel to the magnetic field wherever the field is
not negligible. Due to the low value of the plasma $\beta$ (the ratio of gas pressure to magnetic pressure),
the solar corona is magnetically dominated \cite{Gary}. To describe the equilibrium structure of the static
coronal magnetic field when non-magnetic forces are negligible, the force-free assumption is appropriate:
\begin{equation}
   (\nabla \times\textbf{B})\times\textbf{B}=0 \label{one},
\end{equation}
\begin{equation}
    \nabla \cdot\textbf{B}=0 \label{two},
 \end{equation}
\begin{equation}
    \textbf{B}=\textbf{B}_{\textrm{obs}} \quad \mbox{on photosphere} \label{three},
 \end{equation}
where $\textbf{B}$ is the magnetic field and $\textbf{B}_{\textrm{obs}}$ is measured vector field on the
photosphere. Equation~(\ref{one}) states that the Lorentz force vanishes (as a consequence of
$\textbf{J}\parallel \textbf{B}$, where $\textbf{J}$ is the electric current density) and Equation~(\ref{two})
describes the absence of magnetic monopoles. Based on the above assumption, the coronal magnetic
field is modelled with nonlinear force-free field (NLFF) extrapolation
\cite{Inhester06,valori05,Wiegelmann04,Wheatland04,Wheatland:2009,Wheatland:2011,Amari:2010}.

In this study, we extrapolate the three-dimensional NLFF coronal fields from the photospheric boundary as
successive equilibria. Milne-Eddington inverted vector magnetograms, obtained by the the
\textit{Synoptic Optical Long-term Investigations of the Sun survey} (SOLIS)/\textit{Vector-SpectroMagnetograph}
(VSM), are used as the boundary conditions. From a mathematical point of view, appropriate boundary
condition for force-free modeling are the vertical magnetic field $B_{n}$ and the vertical current $J_{n}$
prescribed only for one polarity of $B_{n}$ \cite{Amari97,Amari99,Amari}. A direct use of these boundary
conditions is implemented in Grad-Rubin codes \cite{Amari99}. Practical computations show, however, that the
solutions for $J_{n}$ described in \inlinecite{DeRosa} $B^{+}_{n}$ and $B^{-}_{n}$ can differ significantly
for real data containing noise and inconsistencies. \inlinecite{Wheatland:2009} and \inlinecite{Wheatland:2011}
implemented a scheme which uses both of $B^{+}_{n}$ and $B^{-}_{n}$ solutions together with an error approximation to
derive a consistent solution. Using the three components of $B$ as boundary condition requires consistent
magnetograms, as outlined in \inlinecite{Aly89}. We use preprocessing and relaxation of the boundary
condition to derive these consistent data on the boundary.

\begin{figure}[htp!]
\begin{center}
   \includegraphics[viewport=10 0 525 375,clip,height=8.6cm,width=11.0cm]{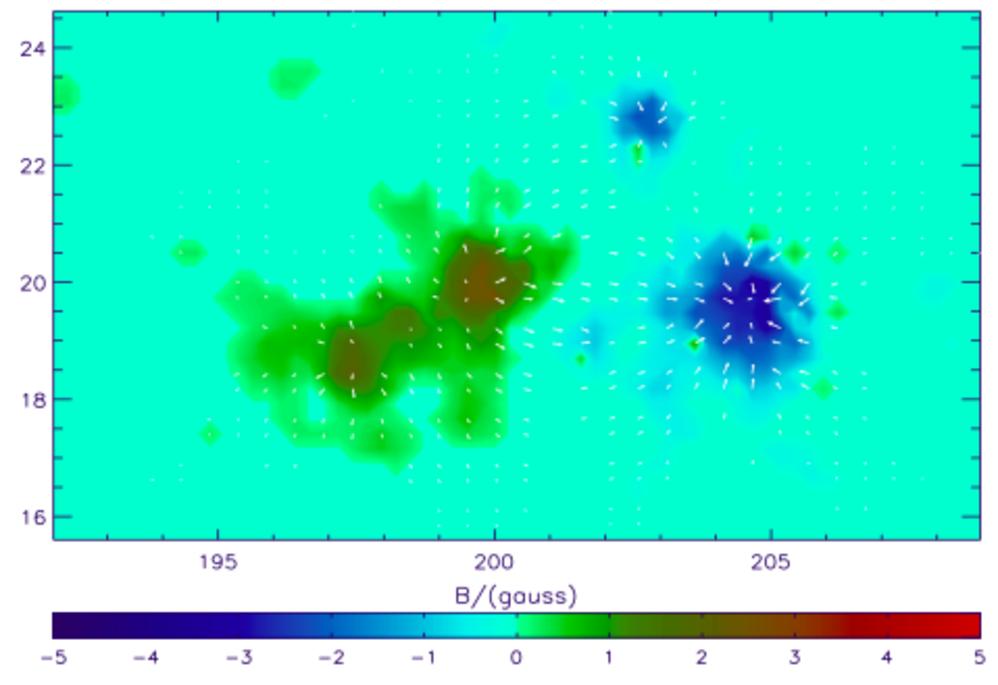}
  \end{center}
     \caption{Surface contour plot of radial magnetic field component and vector field plot of
transverse field( white arrows) of the difference between original data before preprocessing and the final field values
after relaxation through the term $\mathcal{L}_{\textrm{photo}}$ in Equation (\ref{four}). The maximum change
in the transverse field is $30$G which corresponds to the length of the longest white arrow. The vertical and horizontal
axes show latitude, $\theta$(in degree) and longitude, $\phi$(in degree) on the photosphere.
In the area coloured in light blue, field values are lacking. }
\label{figaa1}
\end{figure}

For modeling the coronal magnetic field above the active region and its surrounding, we use the
variational principle originally proposed by \inlinecite{Wheatland00} and later improved by
\inlinecite{Wiegelmann04} in cartesian co-ordinates. The method minimizes a joint measure $(\mathcal{L}_\mathrm{\omega})$
for the normalized Lorentz forces (Equation \ref{one}) and the divergence of the field (Equation \ref{two})
(each of which should equal zero) throughout the volume of interest, $V$. We have implemented this method for
the function $(\mathcal{L}_\mathrm{\omega})$ in spherical geometry \cite{Wiegelmann07,Tilaye:2009}
and iterate $ \textbf{B}$ to minimize $\mathcal{L}_\mathrm{\omega}$. The modification concerns the
input bottom boundary field $\textbf{B}_{\textrm{obs}}$ which the model field $\textbf{B}$ is not
forced to match exactly but we allow deviations of the order of the observational errors. The modified
variational problem is \cite{Wiegelmann10,Tilaye:2010}:
 \begin{displaymath} \textbf{B}=\textrm{argmin}(\mathcal{L}_{\omega}),
\end{displaymath}
 \begin{equation}\mathcal{L}_\mathrm{\omega}=\mathcal{L}_{\textrm{f}}+\mathcal{L}_{\textrm{d}}+\nu\mathcal{L}_{\textrm{photo}}
 \label{four},
\end{equation}
\begin{displaymath} \mathcal{L}_{\textrm{f}}=\int_{V}\omega_{\textrm{f}}(r,\theta,\phi)B^{-2}\big|(\nabla\times\textbf{B})\times
\textbf{B}\big|^2  r^2\sin\theta dr d\theta d\phi,
\end{displaymath}
\begin{displaymath}\mathcal{L}_{\textrm{d}}=\int_{V}\omega_{\textrm{d}}(r,\theta,\phi)\big|\nabla\cdot\textbf{B}\big|^2
  r^2\sin\theta dr d\theta d\phi,
\end{displaymath}
\begin{displaymath}\mathcal{L}_{\textrm{photo}}=\int_{S}\big(\textbf{B}-\textbf{B}_{\textrm{obs}}\big)\cdot\textbf{W}(\theta,\phi)\cdot\big(
\textbf{B}-\textbf{B}_{\textrm{obs}}\big) r^{2}\sin\theta d\theta d\phi,
\end{displaymath}
where $\mathcal{L}_\mathrm{\textrm{f}}$ and $\mathcal{L}_\mathrm{\textrm{d}}$ measure how well
the force-free Equations (\ref{one}) and divergence-free (\ref{two}) conditions are fulfilled,
respectively. $\omega_{\textrm{f}}(r,\theta,\phi)$ and $\omega_{\textrm{d}}(r,\theta,\phi)$ are
weighting functions for the force-free and divergence-free terms, respectively, and are chosen
to be identical for this study. The weighting functions are chosen to be constant within the inner
physical domain $V'$ and decline to 0 with a cosine profile in the buffer boundary region
(see \inlinecite{Tilaye:2009,Tilaye:2010}). The third integral, $\mathcal{L}_{\textrm{photo}}$,
is a surface integral over the photosphere which forces to relax the field $\textbf{B}$ at the
photosphere towards the measured surface field data, $\textbf{B}_{\textrm{obs}}$. In this integral,
$\textbf{W}(\theta,\phi)=\textrm{diag}(w_{\textrm{radial}},w_{\textrm{trans}},w_{\textrm{trans}})$
is a diagonal matrix which gives different weights to the observed surface field components
depending on the  relative accuracy in measurement. In this sense, lacking data is considered most
inaccurate and is taken account of by setting $W(\theta,\phi)$ to zero in all elements of the matrix.
SOLIS/VSM provides full-disk vector-magnetograms, but for some individual pixels the inversion
from line profiles to field values may not have been successful and field data there remains
undetermined at these pixels. Typically, the field is missing where its magnitude is small so that
these pixels have a small impact on the model even if they were measured correctly. Within
the error margin of a measured field value, any value is just as good as any other and from this
range of values we take the value that fits the force-free field best. The different errors for
the radial and transverse components of $\textbf{B}_{\textrm{obs}}$ are taken into account by different
values for $w_{\textrm{radial}}$ and $w_{\textrm{trans}}$. In this work we used $w_{\textrm{radial}}=20w_{\textrm{trans}}$
for the surface preprocessed fields as the radial component of $\textbf{B}_{\textrm{obs}}$
is measured with higher accuracy. Figure~\ref{figaa1} shows the difference in gauss between the
measured vector magnetograms and the final values of the model. Hence the final model field values
on the boundary are consistent with the observed data within noise levels(noise due to fluctuations in intensity)
which are about $1$G and $50$G for longitudinal and transverse components, respectively.
\begin{figure}[htp!]
\begin{center}
   \includegraphics[viewport=15 0 520 340,clip,height=7.80cm,width=11.0cm]{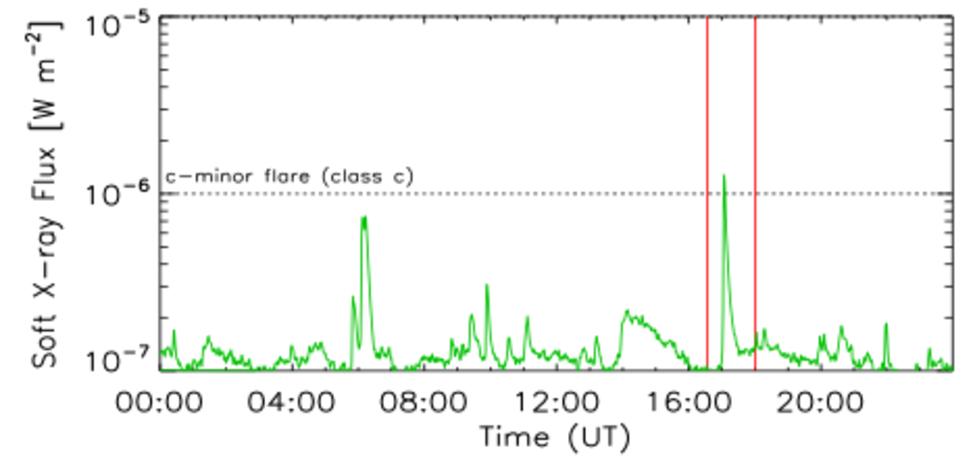}
  \end{center}
     \caption{Solar SXR flux on 27 October 2010 in the wavelength range of
0.1-0.8 nm. Blue vertical lines indicate the availability of SOLIS/VSM data.}
\label{figa}
\end{figure}

We use a spherical grid $r$, $\theta$, $\phi$ with $n_{r}$, $n_{\theta}$, $n_{\phi}$ grid points in the
direction of radius, latitude, and longitude, respectively. For details of the method, we direct the reader
to \inlinecite{Tadesse:2011}. In the present work, we use a larger computational domain which accommodates
most of the connectivity within AR 11117 and its surroundings. We also take the uncertainties of measurements in vector
magnetograms into account as suggested in \inlinecite{DeRosa}.

\section{Results}
\begin{figure}[htp!]
\begin{center}
   \includegraphics[viewport=10 5 445 550,clip,height=16.0cm,width=12.0cm]{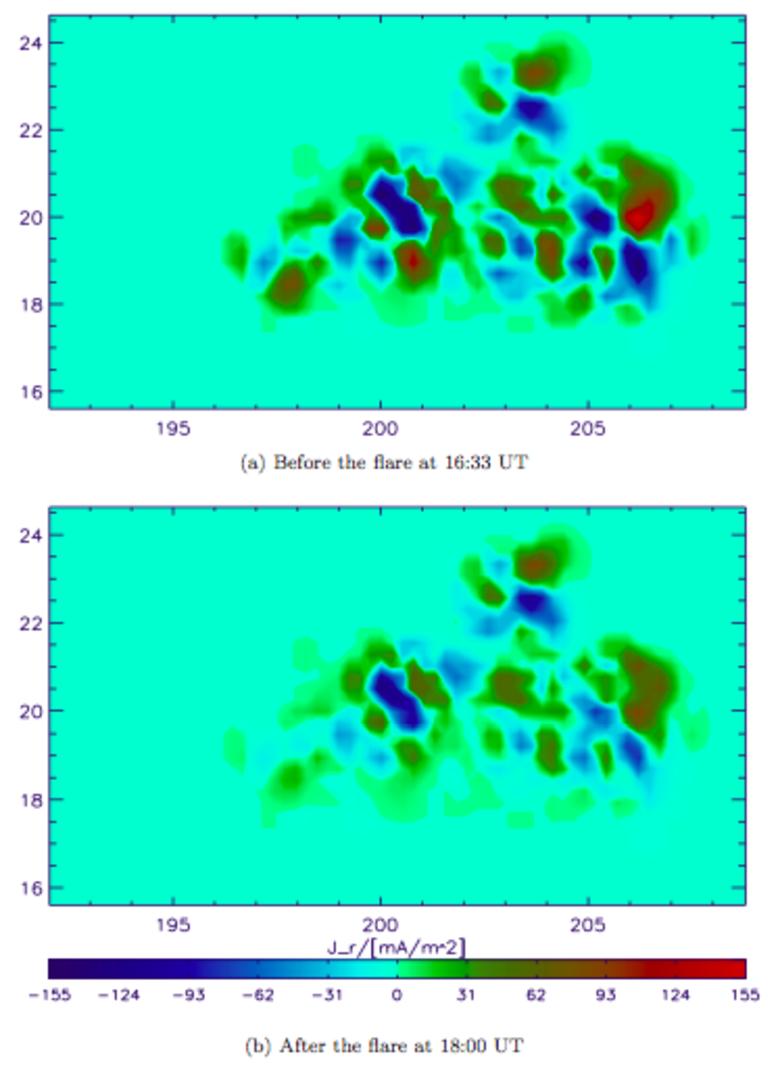}
      \end{center}
     \caption{Surface contour plot of the radial component of electric current density. The color coding shows $J_{r}$ on the photosphere.
     The vertical and horizontal axes show latitude, $\theta$ (in degree) and longitude, $\phi$ (in degree) on the photosphere.
     In the area coloured in light blue, field values are lacking.}
\label{figb1}
\end{figure}

Solar activity on 27 October 2010 was dominated by NOAA AR 11117. A GOES C1.2 flare has been
observed at N$18^0$S$25^0$ in the active region. NOAA records indicate that the
event began in soft X-rays (SXRs) which were detected by the GOES 15 satellite at 16:59 UT, reaching
a peak at 17:04 UT and ending at 17:08 UT (see Figure~\ref{figa}). There were two SOLIS/VSM vector
magnetograms taken, one about half an hour before and the other about one hour after the flare.
In Figure~\ref{figa}, the time the magnetograms were taken is marked by red vertical lines at 16:33 UT
before the flare and the other at 18:00 UT after the flare. As a first step, we remove the
net forces and torques from the boundary using our spherical preprocessing procedure\cite{Tilaye:2009}
which brings the photospheric magnetic field closer to the boundary values of a force-free field
\cite{Molodenskii69,Aly89}. Then we apply our spherical extrapolation scheme using the surface vector
field solution obtained from the preprocessing scheme. To determine the 3D coronal magnetic field
as a \textrm{nlff} equilibrium, we need the three components of the magnetic field of AR 11117 and
its surroundings on the photosphere. The computations are performed in a wedge-shaped
computational box of volume $V$ with $140\times115\times250$ pixels in radial, latitudinal and
longitudinal directions.
\begin{figure}[htp!]
\begin{center}
   \includegraphics[viewport=10 0 445 570,clip,height=16.0cm,width=12.0cm]{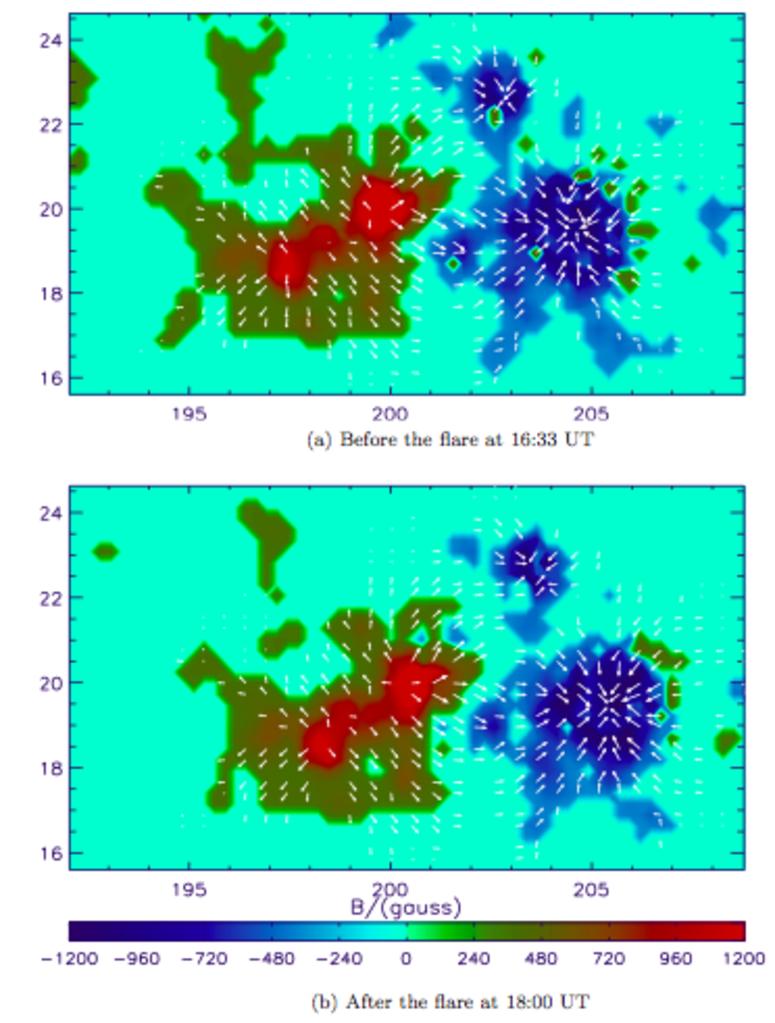}
    \end{center}
     \caption{Surface contour plot of radial magnetic field component and vector field plot of
transverse field with white arrows. The color coding shows $B_{r}$ on the photosphere. The vertical and horizontal
axes show latitude, $\theta$(in degree) and longitude, $\phi$(in degree) on the photosphere. In the area
coloured in light blue, field values are lacking.}
\label{figb}
\end{figure}
The computational box is large enough to include the connectivity between the AR 11117 and its
surroundings. Figure~\ref{figb} shows the temporal variations of the photospheric vector magnetic
components of AR 11117 before and after the flare. The normal component is color coded and the horizontal components
are shown as surface vectors. In order to quantify the change in the surface vector magnetic
field, we computed the vector correlation between the fields before and after the
flare. We use the vector correlation ($C_\mathrm{\textrm{vec}}$) \cite{Schrijver06} metric which
generalizes the standard correlation coefficient for scalar functions and is given by
\begin{equation}
C_\mathrm{ \textrm{vec}}=  \frac{\sum_i \textbf{v}_{i} \cdot \textbf{u}_{i}}{\sqrt{\sum_i |\textbf{v}_{i}|^2} \sqrt{\sum_i
|\textbf{u}_{i}|^2} }\label{nine}
\end{equation}
where $\textbf{v}_{i}$ and $\textbf{u}_{i}$ are 2D vectors at grid point $i$. If the vector
fields are identical, then $C_{\textrm{vec}}=1$; if $\textbf{v}_{i}\perp \textbf{u}_{i}$ , then
$C_{\textrm{vec}}=0$. The correlation ($C_\mathrm{\textrm{vec}}$) of the 2D surface magnetic field
vectors before and after the flare are $0.96$ and $0.87$ for the radial and transverse components, respectively.
From these values we can see that there has been some change in the surface magnetic field
configuration during the flare event. The change in the surface magnetic field
is towards an intensification in transverse components. This change in transverse components
indicates that there is also the change in the vertical components of electric current density.
Figure~\ref{figb1} shows the temporal variation of the vertical electric current density on
the photosphere. We computed the vertical electric current density on the surface using the relation
$J_{r}=\hat{\textbf{r}}\cdot\nabla\times\textbf{B}$ (where \textbf{B} the magnetic field) from the transverse
magnetic field components. We depict a surface color plot of the vertical electric current densities before
and after the flare in Figures~\ref{figb1}a and b, respectively. In order to quantify the change in the radial
electric current densities on the photosphere, we calculate the total of the absolute value of the vertical
electric current density before and after flare. We use the pointwise sum of the values at nodal points.
We find a ratio of the total absolute value of the vertical electric current density after the flare
to that before the flare to be 0.863. The vertical electric current in the active region therefore has
decreased after the flare.

\begin{figure}[htp!]
\begin{center}
\includegraphics[viewport=10 0 545 550,clip,height=13.0cm,width=12.0cm]{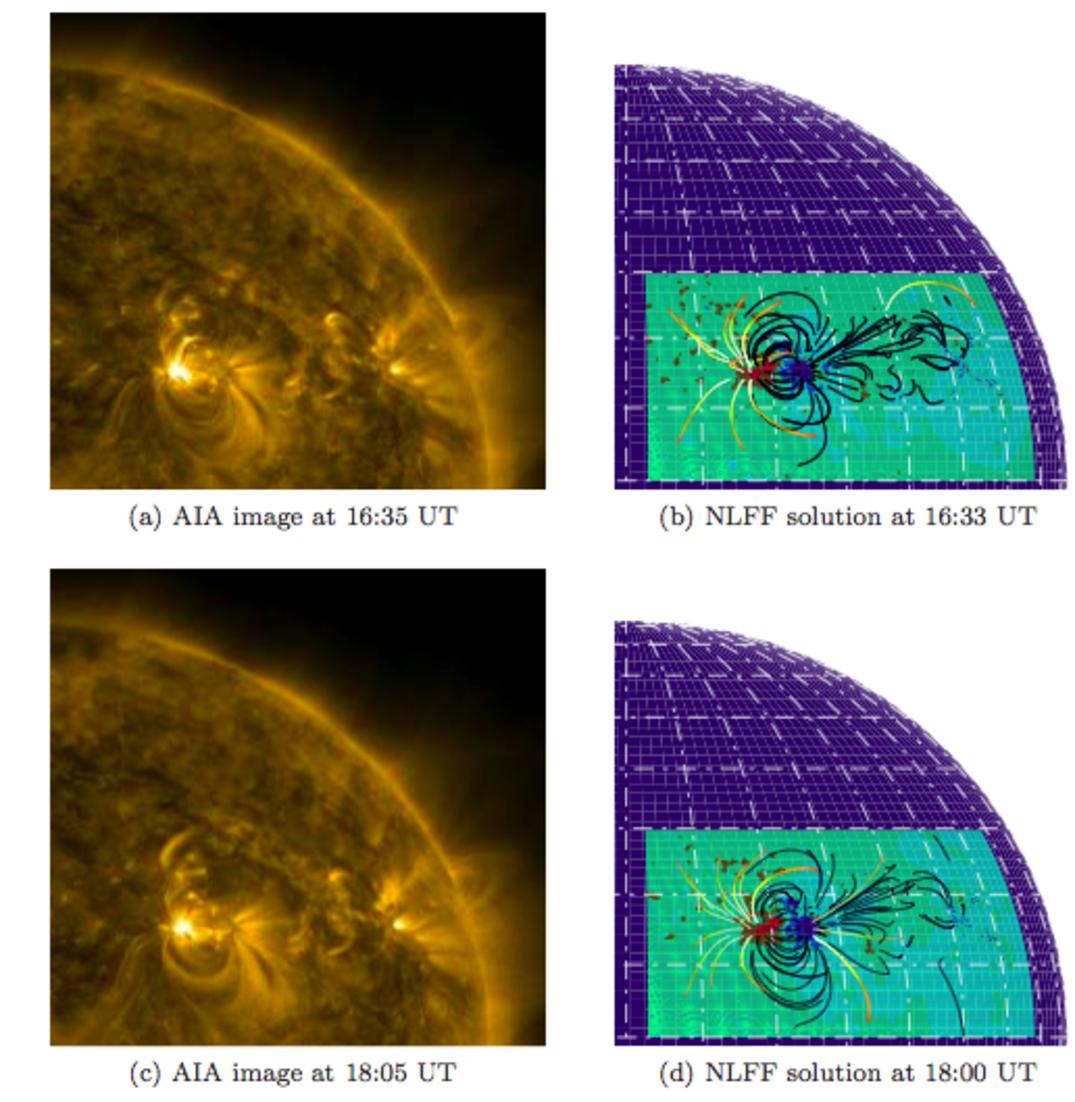}
     \end{center}
     \caption{SDO/AIA ( textit{Solar Dynamics Observatory/ Atmospheric Imaging Assembly}) 171{\AA} images and their respective
     selected magnetic field lines reconstructed from SOLIS magnetograms using nonlinear force-free modelling.}
\label{figc}
\end{figure}
\begin{figure}[htp!]

\begin{center}
   \includegraphics[viewport=10 5 445 550,clip,height=16.0cm,width=12.0cm]{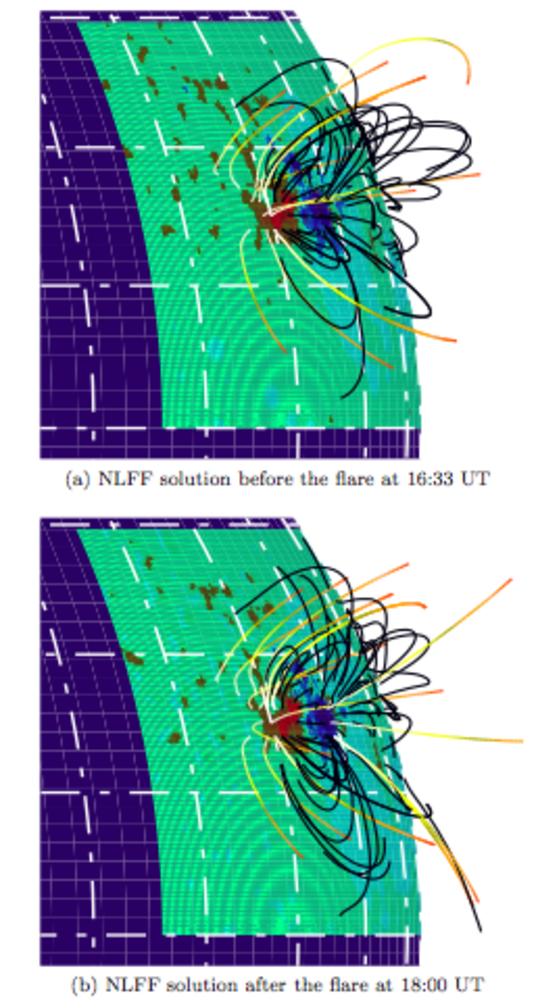}
  \end{center}
     \caption{Some magnetic field lines reconstructed from SOLIS magnetograms. These Figures are obtained by zooming and rotating
     Figures~\ref{figc}b and d to the solar limb. }
\label{fige}
\end{figure}

Magnetic fields are generally recognized as playing a fundamental role in flares. During the course of
a flare, the magnetic field is believed to undergo major changes \cite{Jing:2009}. In this work we have
compared the magnetic field topologies from two datasets taken before and after the flare
using our 3D NLFF reconstruction method. From a visual inspection of the magnetic field lines within
the extrapolation volume, we recognize some changes in the magnetic field structure during the C1.2 flare.
Figure~\ref{figc} shows some selected magnetic field lines from our reconstructions before and after
the flare along with the respective SDO/AIA images. Figure~\ref{fige} shows the magnetic field
lines of Figure~\ref{figc} zoomed in and rotated to the limb. In order to compare the fields at the two
consecutive datasets quantitatively, we computed the vector correlations between the 3D field configurations.
The correlations ($C_\mathrm{\textrm{vec}}$) of the 3D magnetic field vectors before and after the flare
with respect to the potential field configuration before the flare are $0.81$ and $0.93$ respectively.
We see that the magnetic field configuration after the flare looses some of its non-potentiality.

The energy stored in the magnetic field as a result of fieldline stressing into a non-potential
configuration has been identified as the source of flare energy. The study by \inlinecite{Jing:2010}
confirms that there is physical link between magnetic energy and flare occurrence in active regions.
Study of the temporal evolution of the free magnetic energy indicates that it varies before and after
the flare events \cite{Jing:2009}. There is strong need to estimate this free energy numerically.
One way to estimate the energy budget of active regions is to reconstruct the three-dimensional (3D)
coronal field from the measured photospheric boundary based on the force-free assumption. We compute the
free magnetic energy from the excess energy of the extrapolated field beyond that of the potential field
which satisfies the same $\textbf{B}_{\textrm{obs}}\cdot\hat{r}$ boundary condition. Similar estimates have
been made by \inlinecite{Regnier}, \inlinecite{Thalmann}, and \inlinecite{Tadesse:2011} for active regions observed at
other times. From the corresponding potential and force-free magnetic field, $\textbf{B}_{\textrm{pot}}$
and $\textbf{B}$, respectively, we can estimate an upper limit to the free magnetic energy associated with coronal
currents
\begin{equation}
E_\mathrm{free}=E_\mathrm{\textrm{nlff}}-E_\mathrm{\textrm{pot}}=
\frac{1}{8\pi}\int_{V'}\Big(B_{\textrm{nlff}}^{2}-B_{\textrm{pot}}^{2}\Big)r^{2}\textrm{sin}\theta dr d\theta d\phi. \label{ten}
\end{equation}
\begin{center}
\begin{table}
\begin{tabular}{cccc}
 \hline
 Events & $E_{\textrm{nlff}}(10^{32}\textrm{erg})$&$E_\mathrm{\textrm{pot}}(10^{32}\textrm{erg})$& $E_\mathrm{free}(10^{32}\textrm{erg})$\\
\hline
Before the flare at 16:33UT &$12.89$&$10.84$&$2.05$\\
After the flare at 18:00UT &$10.93$&$10.62$&$0.31$\\
\hline
\end{tabular}
\caption{The magnetic energy obtained from potential and NLFF field extrapolations before and after the flare within
the computational box.}
\label{table1}
\end{table}
\end{center}
The computed energy values are listed in Table~\ref{table1}. The change in the free energy during
the flare is about $1.74\times10^{32}\,\textrm{erg}$. The magnetic energy of the potential field
configuration is about $10\times10^{32}\textrm{erg}$. $E_{\textrm{nlff}}$ exceeds
$E_{\textrm{pot}}$ by only 15.9$\%$ and 2.8$\%$ for before and after the flare, respectively.
To estimate the uncertainty in the numerical result, the code was applied to the original SOLIS
data to which artificial random noise had been added in the form of a normal distribution with an
amplitude of approximately 1 \textrm{G} in the longitudinal and 50 \textrm{G} in the transverse component.
The chosen noise amplitudes are based on the sensitivity of the VSM instrument. Hence, we found that the
evaluated relative error of the energy estimation is about 0.4\% for the potential and 1\% for the NLFF
field (i.e. $E_\mathrm{\textrm{pot}}\pm 0.044\times10^{32}\,\textrm{erg}$ and $E_{\textrm{nlff}}\pm 0.129\times10^{32}\,\textrm{erg}$,
respectively). The available free magnetic energy is approximately $10^{32}\,\textrm{erg}$ with
a relative error of about 4\% (i.e. $E_\mathrm{free}\pm 0.082\times10^{32}\,\textrm{erg})$. In principle high spatial
resolution of a magnetogram gives a higher magnetic energy because small-scale magnetic variations
are better resolved. How much better resolution contributes to the total energy estimate depends on
the steepness of the spectral energy distribution. Steeper spectra yield less additional energy if the
resolution is enhanced, flatter spectra yield more additional energy. In addition to
our relative error estimates above, low spatial resolution might increase the
error.

To study the influence of the use of preprocessed boundary data along with the departure of the model
field from the observed boundary fields on the estimation of free-magnetic energy, we have computed the
magnetic energy of the potential field and the NLFFF obtained from the original data without preprocessing
and with preprocessing. As the preprocessing procedure filters out small-scale surface field fluctuations,
the magnetic energy of NLFFF obtained from preprocessed boundary data is smaller than the corresponding
energy without preprocessing. The energy of the potential field obtained from boundary data with and without
preprocessing are close in value, since the potential field calculation makes only the use of the radial
magnetic field component which is not affected too much by the preprocessing. The magnetic energy computed
from the original data without preprocessing is about $13.67\times10^{32}\textrm{erg}$ which is about
$6\%$ higher than the one obtained from preprocessed and modified observational boundary data. However,
this energy does not correspond to the nonlinear force-free magnetic field solution since the original
boundary data without preprocessing is not a consistent boundary condition for NLFFF modeling.
\begin{figure}[htp!]
\begin{center}
   \includegraphics[viewport=10 0 665 470,clip,height=6.80cm,width=12.0cm]{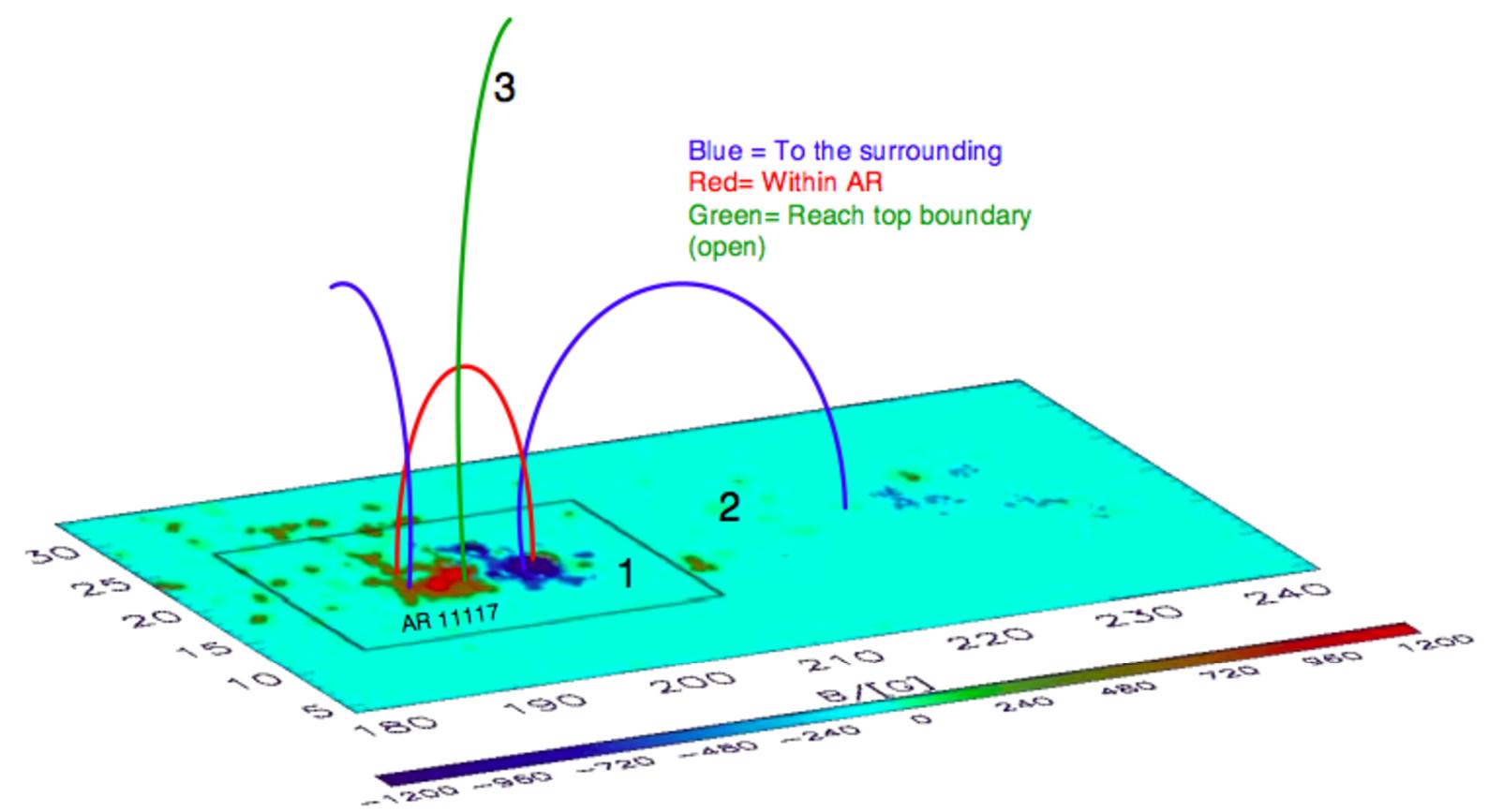}
  \end{center}
     \caption{The connectivity between AR 11117 and its surroundings. The black rectangle shows the domain of AR 11117.
     The color coding shows $B_{r}$ on the photosphere. The red fieldline represents fieldlines connecting opposite
     polarities within the active region. The blue one represents fieldlines connecting the active region with its
     surroundings both on the photosphere and the side boundaries (categorized as elsewhere). The green one shows fieldlines,
     which leave the computational box though its top boundary and may be considered to be "open".}
\label{figg}
\end{figure}

In our previous work \cite{Tadesse:2011}, we have studied the connectivity between three neighbouring active
regions. In that study, we investigated the three ARs were found to share a significant amount of magnetic flux compared
to their internal flux connecting one polarity to the other. In terms of the electric current they were much more
isolated. In this work, we study the connectivity between AR 11117 and its surroundings before and after
flare. In order to quantify these connectivities, we have calculated the magnetic flux and the electric
currents shared between the active region and its surroundings. For the magnetic
flux, \textit{e.g.}, we use
\begin{equation}
\Phi_{\alpha\beta}=\sum_{i}|\textbf{B}_{i}\cdot\hat{r}|R^{2}_{\odot}\textrm{sin}(\theta_{i})\Delta\theta_{i}\Delta\phi_{i}\label{ten1}
\end{equation}
where the summation is over all pixels of $\mbox{AR}_{\alpha}$ from which the field line
ends in $\mbox{AR}_{\beta}$ or $i\in\mbox{AR}_{\alpha}\|\,\mbox{conjugate footpoint}(i)\in\mbox{AR}_{\beta}$.
The indices $\alpha$ and $\beta$ may take values between $1$ and $3$. The index number $1$ corresponds
to AR $11117$, index number $2$ to its surroundings and the side boundaries and index number $3$ to the
top boundary (see Figure~\ref{figg}). For the electric current we replace the magnetic field,
$\textbf{B}_{i}\cdot\hat{r}$, by the vertical current density $\textbf{J}_{i}\cdot\hat{r}$ in
Equation (\ref{ten1}). Both Table~\ref{table2} and~\ref{table3} show the percentage of the total
magnetic flux and electric current shared between the AR 11117 and its surroundings before and after
the flaring event. For example, first column of Table~\ref{table2} shows that $82.23\%$ of positive
polarity of AR 11117 is connected to negative polarity within itself; line 2 shows that $41.51\%$ of
positive/negative polarity of AR 11117 is connected to positive/negative polarity of
its surroundings including the side boundaries of the computational box, and line 3 shows that $74.78\%$ of
the total magnetic flux of top boundary of the computational box connected to the positive/negative polarity
of AR 11117. Table~\ref{table3} shows the electric current connectivity we have calculated applying the same
technique. Figure~\ref{figb1} shows that the vertical electric current density has decreased after the
flare. In this study (see Table~\ref{table3}), we found that AR 11117 is even more isolated in electric
current from its surroundings after the flare. It is noteworthy that modeling this active region with
small cartesian box would lead to wrong NLFF model solution as there are currents crossing
its boundaries.

\begin{table}
\begin{tabular}{rlclrcr}
\cline{2-7}
&& $\textrm{Before the flare}$& &&$\textrm{After the flare}$& \\
\cline{2-7}
$\Phi_{\alpha\beta}$ &$\beta=1$ & $2$ & $3$&$\beta=1$ & $2$ & $3$\\
\cline{1-7}
\multicolumn{1}{r}{$\alpha=1$}& $82.23$&$12.61$&$5.16$&   $88.37$&$7.36$&$4.27$\\
\multicolumn{1}{r}{$2$}& $41.51$&$50.94$&$7.55$&  $35.18$&$56.24$&$8.58$\\
\multicolumn{1}{r}{$3$}& $74.78$&$25.22$&$0.00$&    $65.49$&$34.51$&$0.00$\\
\cline{1-7}
\end{tabular}
\caption{The percentage of the total magnetic flux shared between AR 11117 and its surroundings.
$\Phi_{11}$, $\Phi_{22}$ and $\Phi_{33}$ denote magnetic flux of AR 11117, outside AR 11117 on the
photosphere including the side boundaries and the top boundary of the computational box (see Figure~\ref{figg}),
respectively. }
\label{table2}
\end{table}
\begin{table}
\begin{tabular}{rlclrcr}
\cline{2-7}
&& $\textrm{Before the flare}$& &&$\textrm{After the flare}$& \\
\cline{2-7}
$I_{\alpha\beta}$ &$\beta=1$ & $2$ & $3$&$\beta=1$ & $2$ & $3$\\
\cline{1-7}
\multicolumn{1}{r}{$\alpha=1$}& $93.02$&$6.98$&$0.00$&   $98.06$&$1.94$&$0.00$\\
\multicolumn{1}{r}{$2$}& $34.41$&$65.59$&$0.00$&  $26.38$&$73.62$&$0.00$\\
\multicolumn{1}{r}{$3$}& $0.00$&$0.00$&$0.00$&    $0.00$&$0.00$&$0.00$\\
\cline{1-7}
\end{tabular}
\caption{The percentage of the total electric current shared between AR 11117 and its surroundings.
$I_{11}$, $I_{22}$, and $I_{33}$ denote electric current within AR 11117, outside AR 11117 on the photosphere
including the side boundaries and the top boundary of the computational box (see Figure~\ref{figg}),
respectively.}
\label{table3}
\end{table}
\section{Conclusions}
\label{sect:disc}
We have investigated the coronal magnetic field associated with AR 11117 and
its surroundings observed on 27 October 2010 by analysing SOLIS/VSM data. Two vector
magnetograms with a time cadence of 1 hour and 28 min were available to investigate
the magnetic energy content of the coronal field during the C1.2 flare observed by GOES.
We have used an optimization method for the reconstruction of nonlinear force-free
coronal magnetic fields in spherical geometry by restricting the code to limited
parts of the Sun \cite{Wiegelmann07,Tilaye:2009,Tilaye:2010,Tadesse:2011}.

We have studied the time evolution of the magnetic field from before to after the
flare. We found that there is some rearrangement in the magnetic field configuration
after the eruption. The magnetic energies calculated in a large wedge-shaped
computational box above the active region and its surroundings decreased after the flare,
indicating that the field looses some of its non-potentiality. However, caution is
needed when estimating the free magnetic energy using NLFFF modeling. Many aspects of the
specific approach used in NLFFF modeling may influence the results. This is the
first study which contains AR and its surroundings with a flaring event in our model.
It was made possible by the use of spherical coordinates and allowed us to
analyse connectivity between AR and its surroundings. Modeling an active region
in cartesian geometry would lead to wrong NLFF model solution as there are currents
crossing the small cartesian box enclosing it by excluding its surroundings. We propose
to systematically study the effect of using cartesian box over spherical wedge-shaped
box for NLFF solutions.

High cadence magnetogram observations are necessary when we study the magnetic field
topology and energy variations associated with CME/flare eruptions. In this sense,
it is worth mentioning that the \textit{ Helioseismic and Magnetic Imager} (HMI) on
board the \textit{Solar Dynamic Observatory} (SDO) is the first instrument to provide
routine measurements of the full-disk photospheric vector magnetogram data with high
spatial and temporal resolution under seeing-free condition. We anticipate extending
the current study with soon to be released SDO/HMI full-disk vector magnetograms.

\section*{Acknowledgements} SOLIS/VSM vector magnetograms are produced cooperatively by NSF/NSO and NASA/LWS.
The National Solar Observatory (NSO) is operated by the Association of Universities for Research in Astronomy,
Inc., under cooperative agreement with the National Science Foundation. Tilaye Tadesse Asfaw acknowledges a fellowship of
the International Max-Planck Research School at the Max-Planck Institute for Solar System Research and the work of T.
Wiegelmann was supported by DLR-grant $50$ OC $0501$.

\bibliographystyle{spr-mp-sola}
\bibliography{paper_2011_bibtex}

\end{article}
\end{document}